\documentclass[aps,prb,twocolumn,superscriptaddress,showpacs]{revtex4-2}
\usepackage{amsmath,amsfonts, amssymb}
\usepackage{bm}
\usepackage{graphicx}
\usepackage{verbatim}
\usepackage{hyperref}
\usepackage{xcolor}
\usepackage{enumerate}
\usepackage{tikz}
\usepackage{bbold}
\usepackage{booktabs}
\usepackage{multirow}
\usepackage{ulem}

\usepackage[utf8]{inputenc}
\usepackage[T1]{fontenc}
\usepackage{xcolor}

\usepackage{tabularx}

\usepackage{bm}
\DeclareMathOperator{\Tr}{Tr}

\pdfoutput=1
\usepackage{color}
\definecolor{LinkColor}{rgb}{0.256,0.439,0.588}
\usepackage{hyperref}
\hypersetup{
colorlinks=true,
citecolor=LinkColor,
linkcolor=LinkColor,
urlcolor=LinkColor
}

\usepackage{multirow}

\begin{document}
\title{Extracting Universal Corner Entanglement Entropy during the Quantum Monte Carlo Simulation}
\author{Yuan Da Liao}
\affiliation{Department of Physics and HK Institute of Quantum Science \& Technology, The University of Hong Kong, Pokfulam Road, Hong Kong SAR, China}

\author{Menghan Song}
\affiliation{Department of Physics and HK Institute of Quantum Science \& Technology, The University of Hong Kong, Pokfulam Road, Hong Kong SAR, China}

\author{Jiarui Zhao}
\affiliation{Department of Physics and HK Institute of Quantum Science \& Technology, The University of Hong Kong, Pokfulam Road, Hong Kong SAR, China}

\author{Zi Yang Meng}
\affiliation{Department of Physics and HK Institute of Quantum Science \& Technology, The University of Hong Kong, Pokfulam Road, Hong Kong SAR, China}

\date{\today}

\begin{abstract}
The subleading corner logarithmic corrections in entanglement entropy (EE) are crucial for revealing universal characteristics of the quantum critical points (QCPs), but they are challenging to detect.
Motivated by recent developments in the stable computation of EE in (2+1)D quantum many-body systems, we have developed a new method for directly measuring the corner contribution in EE with less computational cost. The cornerstone of our approach is to measure the subtracted corner entanglement entropy (SCEE) defined as the difference between the EEs of subregions with the same boundary length for smooth and cornered boundaries during the sign-problem free quantum Monte Carlo simulation. Our improved method inherently eliminates not only the area law term of EE but also the subleading log-corrections arising from Goldstone modes, leaving the universal corner contribution as the leading term of SCEE with greatly improved data quality. 
Utilizing this advanced approach, we calculate the SCEE of the bilayer Heisenberg model on both square and honeycomb lattices across their (2+1)D O(3) QCPs with different opening angles on entanglement boundary, and obtain the accurate values of the corresponding universal corner log-coefficients. These findings will encourage further theoretical investigations to access controlled universal information for interacting CFTs at (2+1)D.
\end{abstract}

\maketitle

\section{Introduction}
Entanglement entropy (EE) serves as a crucial measure of entanglement information in quantum states and can reveal the universal characteristics of the underlying systems~\cite{calabreseEntanglement2004,casiniEntanglement2009,eisertColloquium2010,laflorencieQuantum2016}.
In recent years, there has been significant progress from field-theoretical analysis and lattice model numerical simulations in understanding the scaling forms of EE for various strongly correlated quantum states. These include spontaneously symmetry-breaking states~\cite{HelmesEEbilayer2014,metlitski2015entanglement,kulchytskyyDetecting2015,demidioEntanglement2020,zhaoMeasuring2022,dengImproved2023,zhouIncremental2024,songExtracting2023,panStable2023}, gapped topologically ordered states~\cite{isakovTopological2011,BlockKogome2020,zhaoMeasuring2022}, and more related to the present work, $(2+1)$D O$(N)$~\cite{kallinCorner2014,HelmesEEbilayer2014,zhaoScaling2022,songExtracting2023}, deconfined~\cite{zhaoScaling2022,liaoTeaching2023,songExtracting2023,songDeconfined2023,deng2024diagnosing,demidioEntanglement2024,zhaoScaling2024}, Gross-Neveu~\cite{liaoTeaching2023,daliaoControllable2023,demidioUniversal2022} and symmetric mass generation~\cite{liuDisorder2023} quantum critical points (QCPs).
In these studies, the subleading logarithmic correction to the area law of EE plays a pivotal role in demonstrating the underlying universal properties.

\begin{figure}[htp!]
\includegraphics[width=\columnwidth]{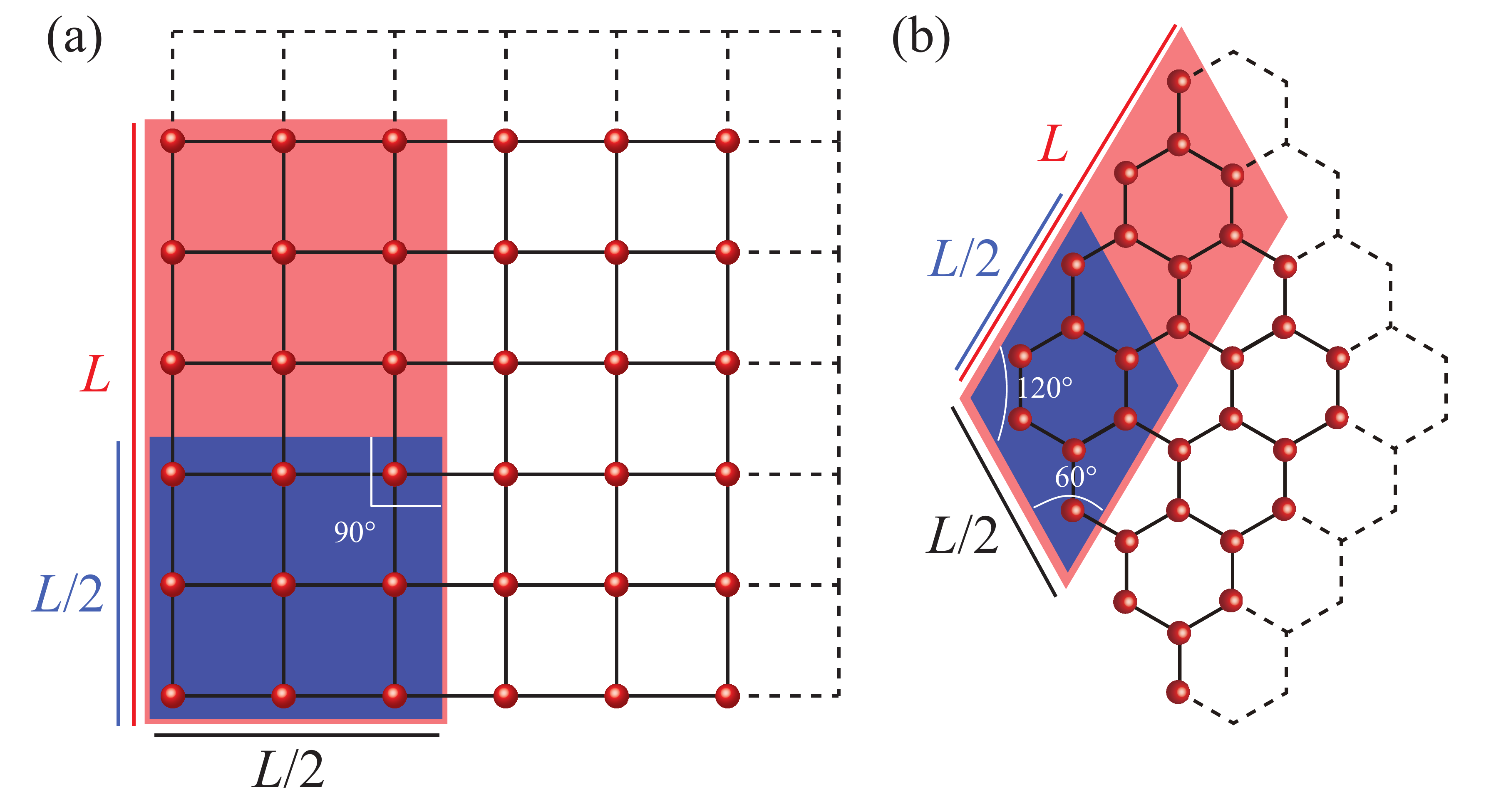}
\caption{\textbf{Illustrations of the entanglement subregions.} Here we use the Heisenberg bilayer (Eq.~\eqref{eq:eq7}) on (a) square lattice and (b) honeycomb lattice to illustrate the entanglement region $A_1$ (red) with smooth boundaries and the entanglement region $A_2$ (blue) with sharp corners. The lattice is periodic in both directions as represented by black dash lines. 
}
\label{fig:fig1}
\end{figure}

To be more specific, the scaling of 2nd order R\'enyi EE at a QCP of a 2D lattice model, described by a conformal field theory (CFT), is given by
~\cite{fradkinEntanglement2006,casiniUniversal2007},
\begin{equation}
S^{(2)}_{A}(l_A) = a l_A- s \ln l_A +c+O(1/l_A),
\label{eq:eq1}
\end{equation}
where $l_A$ is the length of the boundary between the entanglement region $A$ and the environment $\overline{A}$ (usually proportional to the linear system size $L$, as shown in Fig.~\ref{fig:fig1}), $a$ is a non-universal coefficient of the area-law term, 
$s$ is the universal coefficient of the subleading logarithmic correction (log-coefficient), $c$ is a constant, and $O(1/l_A)$ denotes the finite-size correction. 

The universal log-coefficient $s$ depends on the geometry of the entanglement region $A$ for a given unitary CFT~\cite{fradkinEntanglement2006,casiniUniversal2007}. It can be written as $s=\sum_i s(\alpha_i)$, where $\alpha_i$ is the opening angle of the $i$-th corner on the boundary. $s$ satisfies a few universal constraints~\cite{Casini2012,buenoBounds2016,faulknerShape2015,buenoUniversality2015}, including, i). in a CFT $s(\pi)=0$, meaning smooth entanglement cuts (such as the red bipartitions in Fig.~\ref{fig:fig1}) have no log-correction; ii). in a unitary CFT $s(\alpha) > 0$ for $\alpha\in(0,\pi)$; and iii). according to Refs.~\cite{helmesUniversal2016,wucorner2021,Estienne_2022}, the $\alpha$ dependence of $s(\alpha)$ per corner can be evaluated for free Gaussian theories. To the best of our knowledge, besides (1+1)D and free theories~\cite{casiniUniversal2007,casiniEntanglement2009,Casini2012,helmesUniversal2016,helmes2017entanglement}, there exists no rigorous result of the universal corner log-coefficient $s(\alpha)$ at (2+1)D interacting CFTs, even the O(N) CFTs. The precise values of $s(\alpha)$ and its angle-dependence, will provide valuable guidance for the development of controlled analytical theories for (2+1)D CFTs.

In addition, a spontaneous symmetry breaking (SSB) phase with Goldstone modes is expected to exhibit a scaling form of the EE analogous to Eq.~\eqref{eq:eq1}, 
\begin{equation}
S^{(2)}_{A}(l_A) = a l_A- (s_G+s) \ln l_A +c+O(1/l_A),
\label{eq:eq2}
\end{equation}
but with an additional (larger in absolute value) log-coefficient $s_\text{G}=- n_\text{G}/2$ and $n_\text{G}$ corresponds to the number of Goldstone modes~\cite{metlitski2015entanglement}. In case of the 2D square lattice antiferromagnetic model (the N\'eel ground state breaks the SU(2) symmetry and $n_\text{G}=2$), previous QMC studies have revealed that $s_\text{G}=-1$ and $s=0.05(1)$ if the entanglement cut contains four $\alpha=\pi/2$ corners~\cite{demidioEntanglement2020,dengImproved2023,HelmesEEbilayer2014}, consistent with the field theoretical prediction~\cite{metlitski2015entanglement}.

Before this paper, there have been two popular fitting strategies for extracting the universal corner log-coefficient, $s$, from finite-size data of EE. The first approach involves directly fitting the EE for the corner entanglement region according to Eq.~\eqref{eq:eq1}~\cite{demidioEntanglement2020,demidioUniversal2022,panStable2023,liaoTeaching2023,zhaoScaling2022,zhaoMeasuring2022,dengImproved2023,deng2024diagnosing,demidioEntanglement2024,zhouIncremental2024,songDeconfined2023}. This approach is usually used at QCPs since the Goldstone mode contribution $s_G$ should vanished for a CFT, which enables a direct fitting of corner log-coefficient from Eq.~\eqref{eq:eq1}.
The second approach performs a subtraction between the EE obtained from two independent simulations $S_s(l_A) = S_{\text{smooth}}(l_A)-S_{\text{corner}}(l_A)$ -- one with a smooth boundary and one with a corner -- to cancel out the dominant area law contributions and isolate the subtracted corner entanglement entropy (SCEE) with $s \ln l_A$ term as the leading contribution~\cite{levinDetecting2006,kitaevTopological2006,isakovTopological2011,HelmesEEbilayer2014,wangScaling2021,zhouIncremental2024,songExtracting2023,zhaoScaling2024}. However, both of these conventional methods face significant challenges.

For the first approach, extracting the universal subleading log-coefficient $s$ from Eq.~\eqref{eq:eq1} is challenging as the dominant area law term itself exhibits substantial finite-size effects and computational complexity increases considerably with growing system size. Furthermore, in the spontaneous symmetry-breaking (SSB) phase, the log-coefficient originates from both Goldstone modes $s_G$ and corner effects $s$. However, the former is typically much larger than the latter and displays strong finite-size effects~\cite{dengImproved2023}, rendering the isolation of $s$ from $s_G$ unattainable. For a usual QCP, though $s_G$ vanishes at the thermal dynamic limit, it may still appear at finite sizes~\cite{songExtracting2023}, rendering an unclean isolation of $s$ from $s_G$ with small system sizes.
The second subtraction-based approach to obtain SCEE, on the other hand, requires computing the independent EEs for both smooth and corner entanglement regions, which at least doubles the computational costs and leads to worse precision due to error accumulation from the subtraction. 

Moreover, in cases where the nature of QCP is ambiguous (weakly first-order and/or exhibits coexisting finite moments from Goldstone modes) - as observed in recent finite-size analyses of EE at deconfined QCPs~\cite{zhaoScaling2022,songDeconfined2023,songExtracting2023,liaoTeaching2023,deng2024diagnosing,demidioEntanglement2024,zhaoScaling2024} -the stable extraction of universal log-coefficient becomes more challenging yet crucial, as the finite-size effect for the EE becomes more severe, and there exits a log-corrections even for smooth entanglement boundary whose nature is still not well understood.

In light of the above challenges, we develop a new method for spin/bosonic systems to calculate the SCEE {\it by canceling the leading area-law contribution during the Monte Carlo simulation}. This method offers more than twice the reduction in computational cost and much improved numerical accuracy compared to the traditional subtraction-based approach.
In addition to automatically canceling out the area-law term, the SCEE method also cancels the contribution from Goldstone modes, exposing the universal corner contributions as the leading term for both the QCP and SSB states.
With this superior method, we calculate the SCEE of the bilayer Heisenberg model on both square and honeycomb lattices across their (2+1)D O(3) QCPs with different opening angles for corners and obtain the accurate values of the corresponding universal log-coefficients. These findings, {both the more precise universal corner log-coefficient at a strongly interaction (2+1)d QCP and its angle dependence, will provide the benchmark results for the development of controlled analytical theories for strongly interacting (2+1)d CFTs.}

\section{SCEE Methodology}
The SCEE method is designed for directly calculating the subleading universal term of EE arising from corners on the entanglement boundary.
One can choose two different entanglement regions, $A_1$ and $A_2$, as illustrated in Fig.~\ref{fig:fig1} (a) and (b), both having the same boundary length $l_{A_1}=l_{A_2}=2L$. In general, for a QCP with a unitary CFT description or SSB states with Goldstone modes, EE in $A_1$ with a smooth boundary scales as,
\begin{equation}
	S_{A_1}^{(2)} = al_{A_1}-s_G \ln l_{A_1}+\gamma_{1} ,
\end{equation}
with $s_G=0$ for a QCP and $s_G = -\frac{N_G}{2}$ for SSB states with Goldstone modes. If the boundary has corners, then
\begin{equation}
	S_{A_2}^{(2)} = al_{A_2}- (s+s_G) \ln l_{A_2} + \gamma_{2}.
\end{equation}
With $l_{A_1}=l_{A_2}=2L$, the SCEE is given as
\begin{equation}\label{eq:Ss}
	S_s =S_{A_1}^{(2)}-S_{A_2}^{(2)}=s\ln L+\gamma,
\end{equation} 
where $\gamma$ is a constant.\\
It has been pointed out that EE is an exponential observable~\cite{daliaoControllable2023, zhangIntegral2023,zhouIncremental2024}, thus directly evaluating the EE with standard QMC simulations would encounter significant convergence issues.
To address these convergence challenges, {a series of} incremental methods have been developed for both fermionic~\cite{daliaoControllable2023,panStable2023,demidioUniversal2022} and spin/bosonic systems~\cite{zhouIncremental2024}. Additionally, the non-equilibrium method~\cite{albaOut2017,demidioEntanglement2020} can also help mitigate the convergence issues and obtain reliable EE estimates for spin/bosonic systems. 
In this letter, we combine these incremental and non-equilibrium approaches with our new method for measuring SCEE to precisely obtaining universal corner contributions. 
In the following, we will briefly discuss the implementation of such approaches for calculating $S_s$, with the detailed implementations provided in the SM~\cite{supp}.

The 2nd R\'enyi EE can be given as 
$
e^{-S_A^{(2)}} = Z^{(2)}_{R=A}/ Z^{(2)}_{R=\varnothing}
$, where $Z^{(2)}_{R}$ represents the entanglement partition function.
We could express
\begin{equation}\label{eq:SCE}
e^{-S_s} =e^{-\left(S_{A_1}^{(2)}-S_{A_2}^{(2)}\right)}  = Z^{(2)}_{R=A_1}/ Z^{(2)}_{R=A_2}.
\end{equation}
For incremental SWAP operator method~\cite{zhouIncremental2024} based on projector QMC, we have
\begin{equation}
Z^{(2)}_R = \sum\limits_{l,r} w_{l} w_{r}\left\langle V_{l}^0\left|(-H)^{m/2} \text{SWAP}_R   (-H)^{m/2}\right| V_r^0 \right\rangle,
\end{equation}
where $\left|V_l^0\right\rangle$,$\left|V_r^0\right\rangle$ are arbitrary VB states with Kronecker product of two independent replicas of Hamiltonian, $\sum_l w_l\left|V_l^0\right\rangle$ and $\sum_r w_r\left|V_r^0\right\rangle$ are regarded as trail wave functions suggested in Refs.~\cite{liangSome1988,sandvikLoop2010} with higher projector performance to access real ground properties, $m$ is the projection length, and we set $m = 50N$ with $N=2L^2$ as the number of lattice sites,  and $H$ is the Hamiltonian of the system to be studied. 
The SCEE $S_s$ can be given more explicitly as
\begin{equation}
e^{-S_s} \equiv \frac{\sum\limits_{l,r} \bm{W}_{l,r,A_2} \bm{O}_{l,r,A_1,A_2}}{\sum\limits_{l,r} \bm{W}_{l,r,A_2}},
\end{equation}
where 
\begin{equation}
\bm{W}_{l,r,A_2} = w_{l} w_{r}\left\langle V_{l}^0\left|(-H)^{m/2} \text{SWAP}_{A_2}   (-H)^{m/2}\right| V_r^0 \right\rangle
\end{equation} 
and  
\begin{equation}
\bm{O}_{l,r,A_1,A_2} = \frac{\left\langle V_{l}^0\left|(-H)^{m/2} \text{SWAP}_{A_1}   (-H)^{m/2}\right| V_r^0 \right\rangle }{\left\langle V_{l}^0\left|(-H)^{m/2} \text{SWAP}_{A_2}   (-H)^{m/2}\right| V_r^0 \right\rangle
}. 
\end{equation}
Now, we can compute $S_s$ using {\it the QMC simulation without subtracting afterwards} with the new updating weight $\bm{W}$ and improved sampling observables $\bm{O}$. 

Similarly, for stochastic series expansion (SSE) QMC with a non-equilibrium method at finite temperature~\cite{demidioEntanglement2020,zhaoMeasuring2022}, the 2nd R\'enyi entropy can also be expressed as $
e^{-S_A^{(2)}} = Z^{(2)}_{R=A}/ Z^{(2)}_{R=\varnothing}
$, and Eq.~\eqref{eq:SCE} is valid as well, where now  $Z^{(2)}_{R}$ in configuration space is two replicas with region $R$ glued together in imaginary time. To calculate the ratio of two entanglement partition functions, the two partition functions are quenched slowly from $Z^{(2)}_{R=A_2}$ to $Z^{(2)}_{R=A_1}$ and their ratio is related with the total work done along the tuning process. 
Thus the $S_s$ can be obtained directly from {\it one single tuning process. }
More explicitly, 
$
S_s = - \int_{0}^{1} d\lambda \partial \ln \mathcal{Z}_{A_{1}-A_{2}}^{(2)}(\lambda)/\partial\lambda,
$
the entanglement partition function we are considering here is defined as
\begin{equation}
    \mathcal{Z}_{A_{1}-A_{2}}^{(2)}(\lambda)=\sum_{B \subseteq A_{1}-A_{2}} \lambda^{N_B}(1-\lambda)^{N_{A_{1}}-N_{A_{2}}-N_B} Z_{B+A_2}^{(2)},
\end{equation}
where $Z_{B+A_2}^{(2)}$ in configuration space is two replicas with region $B$ and $A_2$ glued together in imaginary time. 
The details about the implementation of the non-equilibrium method can be referred to Refs.~\cite{demidioEntanglement2020,zhaoMeasuring2022}. We set the total number of quench steps as $5\times 10^7$ for all $L$ in this paper.

\begin{figure}[htp!]
\includegraphics[width=1\columnwidth]{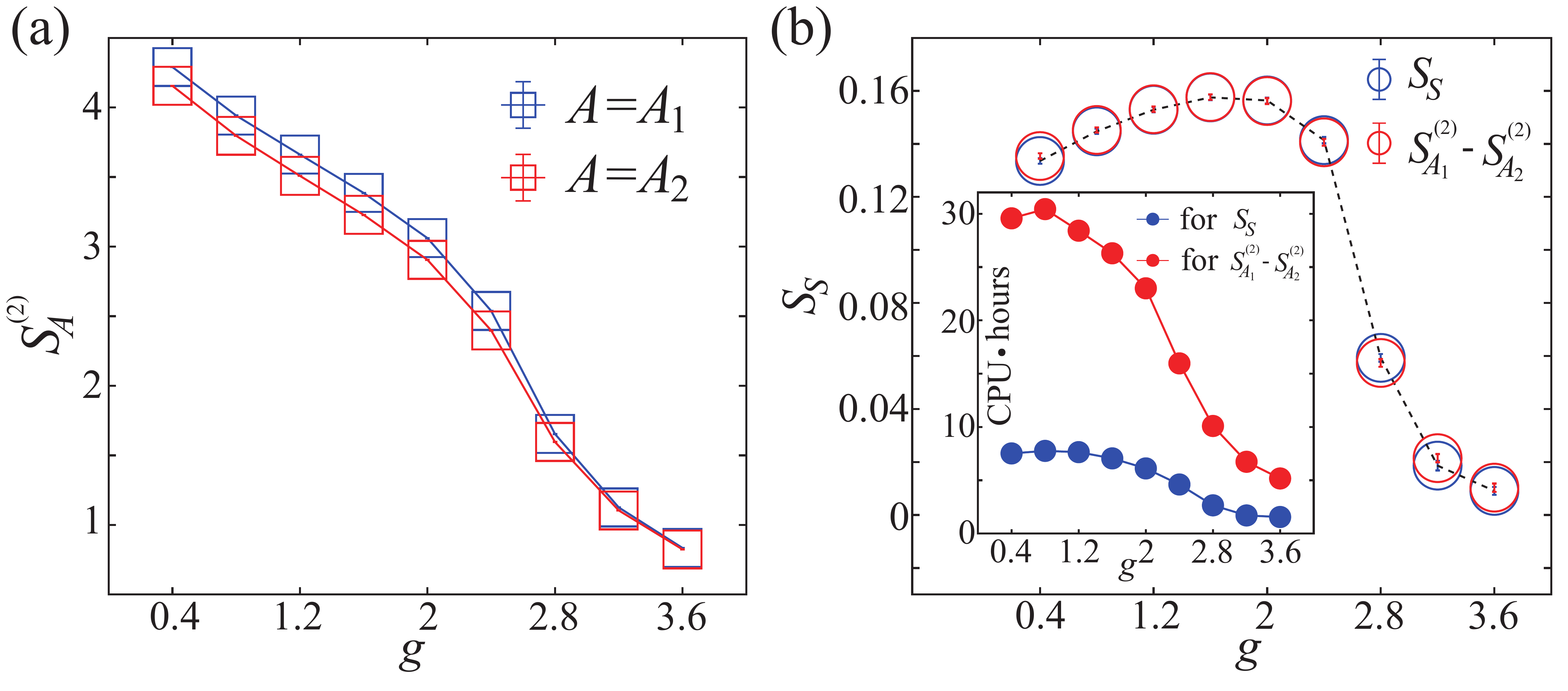}
\caption{\textbf{The superior computational efficiency of SCEE.} We compare the total computational time between $S_s$ and $S_{A_1}^{(2)}-S_{A_2}^{(2)}$ for different $g$ values across the O(3) transition in the square lattice bilayer Heisenberg model with $L=10$. (a) The total computational time is proportional to the total incremental steps $n$. Here, we show that $n$ for $S_s$ is significantly less than $n$ for $S_A^{(1)}$ or $S_A^{(2)}$, which implies that the total computational time for obtaining $S_s$ with the SCEE method is less than half compared to obtaining $S_{A_1}^{(2)}-S_{A_2}^{(2)}$. (b) The values of $S_s$ and $S_{A_1}^{(2)}-S_{A_2}^{(2)}$ are consistent within the error bars for different $g$ values. }
\label{fig:fig2}
\end{figure}

\section{Results}
We apply the SCEE method to the antiferromagnetic Heisenberg bilayer model, both on square and honeycomb lattices, where the $(2+1)$D O(3) phase transition is realized~\cite{sandvikOrder1994,wang2006bilayer,oitmaaGround2012,HelmesEEbilayer2014,songQuantum2023}. As shown in Fig.~\ref{fig:fig1} (a) and (b), the purpose of having both lattices is that we can study the universal corner contributions of EE with different corner opening angles, i.e, in the square case, we have four $\frac{\pi}{2}$ corners and in the honeycomb case two $\frac{\pi}{3}$ and two $\frac{2\pi}{3}$ corners. According to the previous calculations of the Gaussian theory, the angle dependence of the universal corner $s(\alpha)$ can be evaluated for scalar bosons~\cite{helmesUniversal2016}. The $(2+1)$D O(3) transitions are interacting fixed points and we could only take the Gaussian values as qualitative guidelines. Thus extracting the precise values of log-coefficient $s$ of EE at (2+1)D O(3) QCP with numerical tools is a non-trivial task. And we indeed find that the obtained $s(\alpha)$ for $\frac{\pi}{2}$ and $\frac{\pi}{3}$ ($\frac{2\pi}{3}$) are closer to the analytical estimations according to Ref.~\cite{helmesUniversal2016} inside the N\'eel phase where the spin wave is a good description of the ground state.

The antiferromagnetic bilayer Heisenberg model is a system on a bilayer 2D lattice with nearest-neighbor antiferromagnetic intra-layer coupling $J$ and inter-layer coupling $J_{\perp}$, as shown in Fig.~\ref{fig:fig1}. The Hamiltonian is
\begin{equation}
H =J\sum_{\langle i,j \rangle}(\mathbf{S}_{i,1}\cdot\mathbf{S}_{j,1}
		+\mathbf{S}_{i,2}\cdot\mathbf{S}_{j,2})
		+ J_{\perp}\sum_i \mathbf{S}_{i,1}\cdot\mathbf{S}_{i,2},
\label{eq:eq7}
\end{equation}
where $\langle i,j \rangle$ denote the nearest neighbor bonds. We choose $g=J_{\perp}/J$ as the tuning parameter, and previous studies have shown that on square lattice the critical point $g_{c}=2.5220(1)$ separates the N\'eel ordered phase from the inter-layer dimer product phase (i.e. the symmetric phase), and this transition belongs to the (2+1)D O(3) universality class~\cite{wang2006bilayer,lohoeferDynamical2015,wangScaling2022}. 

\begin{figure}[htp!]
\includegraphics[width=\columnwidth]{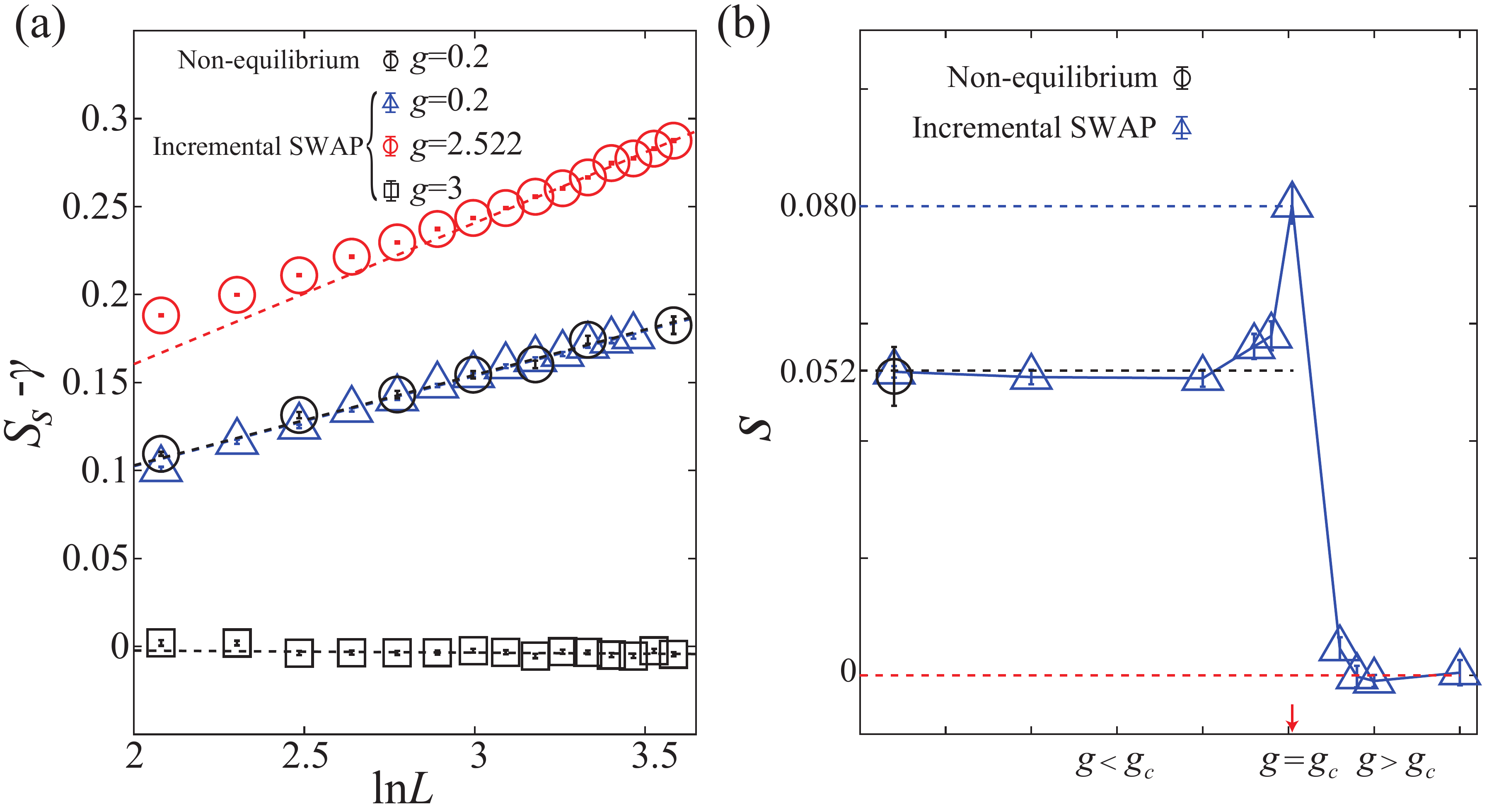}
\caption{\textbf{The $S_s$ for 2D bilayer Heisenberg model on square lattice with four $\pi/2$ corners.} 
(a) The fit of Eq.~\eqref{eq:Ss} with $S_s$ for different $g$ values to obtain the universal coefficient $s$ within the fitting window $\left[L_{\text{min}},L_{\text{max}} \right]$. Here, $L_{\text{min}}=16$ for $g\neq g_c$ and $L_{\text{min}}=24$ for $g= g_c$. The $L_{\text{max}}=36$ corresponds to the largest system size shown in the plot. 
We show the raw data of $S_s$ and also demonstrate the convergence of $s$ as increasing $L_{\text{min}}$ in the SM~\cite{supp}.
(b) The $s \sim 0.052$ for $g<g_c$ is consistent with the reference value of the corner correction inside the N\'eel phase with Goldstone mode~\cite{helmesUniversal2016,metlitski2015entanglement,laflorencieSpin-wave2015}, and the $s=0.080(3)$ at $g_c$ is consistent with the reference value of the corner correct at the O(3) CFT~\cite{kallinCorner2014,HelmesEEbilayer2014,zhaoScaling2022,songExtracting2023}. And for $g>g_c$, the system is inside the gapped dimerized phase, and the corner correction vanishes.}
\label{fig:fig3}
\end{figure}

\begin{figure}[htp!]
\includegraphics[width=\columnwidth]{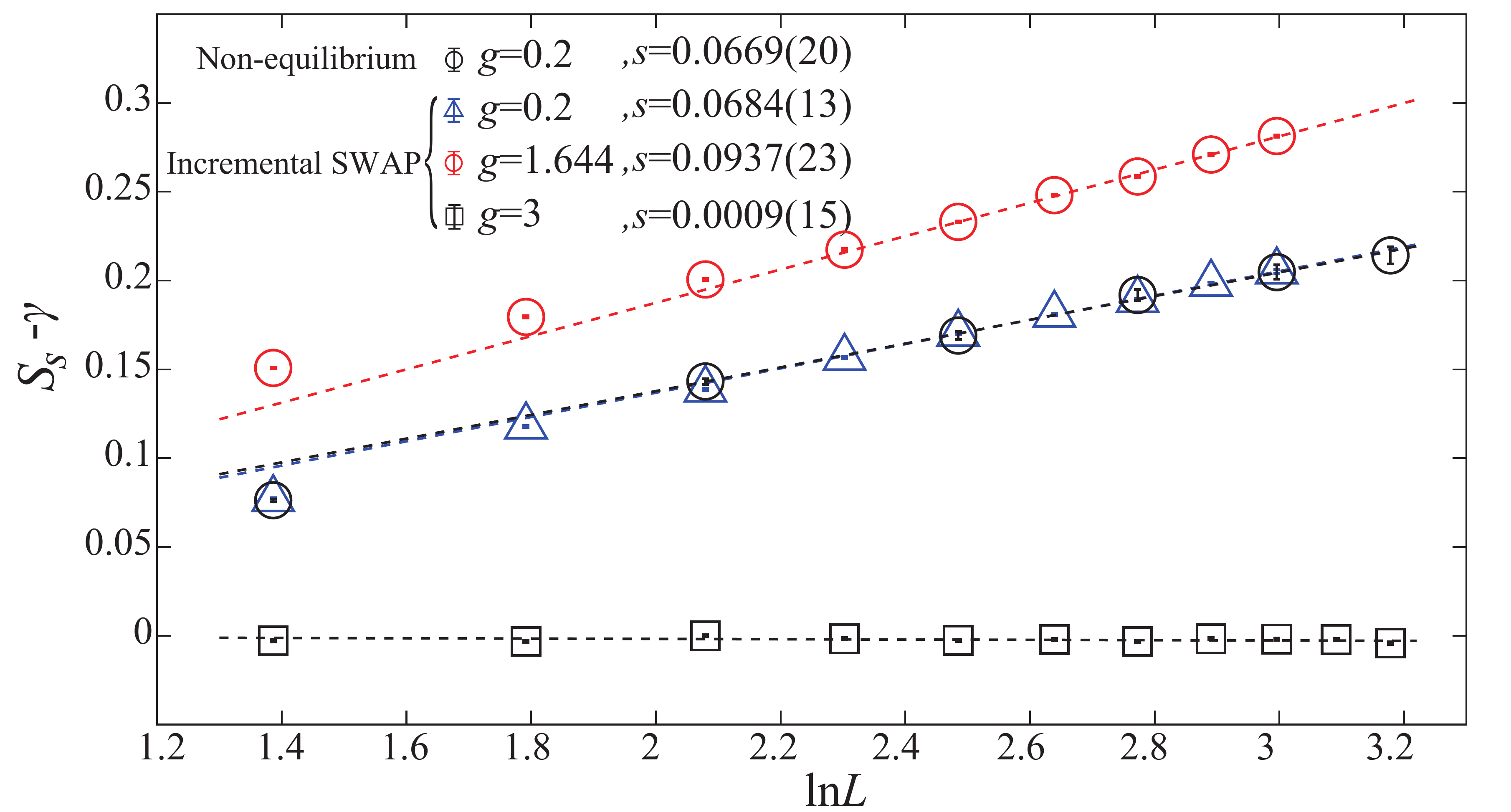}
\caption{\textbf{The $S_s$ for 2D bilayer Heisenberg model on honeycomb lattice with two $\pi/3$ and two $2\pi/3$ corners.} The fit of Eq.~\eqref{eq:Ss} is applied for various $g$ values to obtain $s=2s(\pi/3)+2s(2\pi/3)$ within the fitting window $\left[L_{\text{min}},L_{\text{max}} \right]$, where $L_{\text{min}}=12$ for all $g$ and $L_{\text{max}}=24$ corresponds to the largest system size plotted. For $g=0.2$, a comparison is made between the $S_s$ values obtained using non-equilibrium and incremental SWAP operator methods, resulting in consistent universal coefficients $s$ within the error bars. The values of $s=0.0684(13)$ and $0.0669(20)$ are based on incremental SWAP
operator and non-equilibrium methods, respectively, consistent with theoretical predictions. For $g=3$, $s\sim 0$ as expected. Meanwhile, for $g=g_c$, we obtain $s=0.0937(23)$.
}
\label{fig:fig4}
\end{figure}

Before presenting the $S_s$ at the QCP, we would like to discuss the efficiency of our SCEE algorithm.
Within the incremental SWAP operator method~\cite{zhouIncremental2024}, the computational complexity for a single incremental process is roughly the same, regardless of whether it is used to calculate $S_{A_1}^{(2)}$, $S_{A_2}^{(2)}$, or $S_s$. The total computational effort required to obtain reliable EE is proportional to the total number $n$ of incremental steps for a specific system size $L$ and the parameter $g$.
In the original paper of incremental SWAP operator method~\cite{zhouIncremental2024}, we point out that the total number of incremental steps can be set as the ceiling integer of $\mu$, where $\mu$ is the statistical mean of $\ln\text{SWAP}$. 
Based on this result, we demonstrate that the required $n$ for calculating $S_s$ is significantly lower than the $n$ for $S_{A_1}^{(2)}$ or $S_{A_2}^{(2)}$, as illustrated in Fig.~\ref{fig:fig2} (a). This indicates that the SCEE method reduces the computational cost by more than half, making it highly efficient.
In Fig.~\ref{fig:fig2} (b), we compare the $S_s$ obtained with SCEE method directly and with the subtraction between $S_{A_1}^{(2)}$ and $S_{A_2}^{(2)}$. The values of the $S_s$ for different $g$ are consistent within the error bar. 

Fig.~\ref{fig:fig3} shows our results using the SCEE method for the universal corner corrections. Here, the results are $s=0.051(1)$ and $s=0.051(5)$ at $g=0.2$ based on the incremental SWAP operator and non-equilibrium methods, respectively. These values are in good agreement with previous field-theoretical and numerical results $s=0.052$ for the N\'eel state~\cite{helmesUniversal2016,zhouIncremental2024,HelmesEEbilayer2014}. More importantly, at the QCP of the (2+1)D O(3) transition, the obtained value of $s=0.080(3)$ is fully consistent with previous numerical results~\cite{zhaoScaling2022,songExtracting2023} but with higher precision, and is also consistent with the free Gaussian value within our errorbar. {The corresponding values from free Gaussian theory, in our unit with four $\pi/2$ corners, is 0.078~\cite{casiniEntanglement2009,helmesUniversal2016}}.
We also note that although this corner correction is a universal property of the O(3) CFT, it originates from an interacting fixed point and, therefore, cannot be computed analytically. Our result thus provides the most precise value of this universal quantity obtained so far.  Above the transition, when $g>g_c$, since the Dimer phase has a bulk gap, the entanglement entropy will only have an area-law term and a constant term. Therefore, the corner contribution vanishes, and our obtained value of the log-coefficient is zero, as expected.

For the honeycomb model, we first need to determine the precise position of the QCP, and this is obtained by performing a finite-size scaling analysis of the N\'eel order parameter across the transition using the (2+1)D O(3) critical exponents. As shown in Sec.\uppercase\expandafter{\romannumeral5} in the SM~\cite{supp}, with system sizes $L=12,24,36,48,60,72$, we find that $g_c=1.644(1)$ is the QCP, consistent with the previous  result~\cite{oitmaaGround2012,wang2006bilayer}.

Fig.~\ref{fig:fig4} shows the $S_s$ for bilayer Heisenberg model on honeycomb lattice obtained by SCEE method inside the N\'eel phase ($g=0.2$), at the O(3) QCP ($g=g_c$) and inside the Dimer phase ($g=3$). Similar to the square lattice case in Fig.~\ref{fig:fig3}, the obtained $S_s$ is linearly proportional to $\ln L$, which means we have successfully removed the area law contribution to the EE. Moreover, we obtained the $s=2\times(s(\frac{\pi}{3})+s(\frac{2\pi}{3}))$ are $0.0684(13)$ and $0.0669(20)$ based on incremental SWAP operator and non-equilibrium methods, respectively, inside the N\'eel phase. These values are consistent with the free theory estimation of $~0.06880189$ according to Ref.~\cite{helmesUniversal2016}. 
At the O(3) QCP, we get $s=0.0937(23)$. The obtained $s$ ratio at the QCPs between honeycomb and square lattices, given by $\frac{2s(\frac{\pi}{3})+2s(\frac{2\pi}{3}))}{4s(\frac{\pi}{2})}$, is $1.17(5)$, which is lower than the ratio for free bosons, $1.3231$, {again signifying the difference between the interacting CFT and the free ones.}


\section{Discussion}
In this work, by realizing the EE and SCEE as exponential observables~\cite{daliaoControllable2023,zhangIntegral2023} and employing the incremental procedure~\cite{zhouIncremental2024} that properly handles the smooth boundary and corner cuts in the entanglement partitions, we demonstrate that one can indeed directly obtain the universal corner correction to the EE {during the QMC simulation}
and completely remove the other contributions while sampling the exponential observable.
We showcase the power of the SCEE algorithm in 2D bilayer Heisenberg spin models on square and honeycomb lattices, with the universal corner log-coefficient across their corresponding QCPs being determined. To our knowledge, the angle dependence of the logarithmic coefficient is obtained for the first time in 2D interacting quantum many-body systems. Countless new quantum critical points in 2D highly entangled quantum matter are awaiting exploration.

{\it Note added}: Recently, we become aware of the fundamental concept of subtracted corner entanglement entropy was originally presented in Dr. Helmes’ PhD dissertation~\cite{helmes2017entanglement}.

\begin{acknowledgements}
\noindent{\textcolor{blue}{\it Acknowledgements.}---} 
We acknowledge the inspiring discussions with Michael Scherer, Meng Cheng, Lukas Jassen, and Cenke Xu on the scaling of EE on related projects~\cite{songExtracting2023,songDeconfined2023}. Y.D.L., M.H.S. J.R.Z. and Z.Y.M. acknowledge the support from the Research Grants Council (RGC) of Hong Kong Special Administrative Region (SAR) of China (Project Nos. 17301721, AoE/P701/20, 17309822, C7037-22GF, 17302223), the ANR/RGC Joint Research Scheme sponsored by RGC of Hong Kong and French National Research Agency (Project No. A HKU703/22) and the HKU Seed Funding for Strategic Interdisciplinary Research. Y.D.L also acknowledges the support from the URC Seed Fund for Basic Research for New Staff. We thank HPC2021 system under the Information Technology Services, University of Hong Kong, as well as the Beijng PARATERA Tech CO.,Ltd. (URL:https://cloud.paratera.com) for providing HPC resources that have contributed to the research results reported
within this paper.
\end{acknowledgements}

\bibliography{ref.bib}

\begin{thebibliography}{58}
\bibitem [{sup()}]{supp}%
  \BibitemOpen
  \href@noop {} {\bibinfo {title} {In appendix a, we provide a detailed
  derivation of the entanglement entropy as an exponential observable and then
  discuss how to convert the expectation value of an exponential quantity into
  standard expectation value without exponential in the incremental algorithm.
  in appendix b and c, we provide detailed derivation of the scee in the
  incremental swap scheme with more data. in appendix d we compare the scee
  data between incremental swap and non-equilibrium scheme, as well as the
  standard and titled cuts of the entanglement boundary. in appendix e, we
  present the qmc results for determining the precise qcp for the bilayer
  heisenberg model on honeycomb lattice. in appendix f, we provide the raw data
  of $s_s$ for 2d bilayer heisenberg model.}}\BibitemShut {Stop}%
\end{thebibliography}
\bibliographystyle{apsrev4-2}

\clearpage
\onecolumngrid

\begin{center}
	\textbf{\large Supplemental Material for \\"Extracting Universal Corner Entanglement Entropy during the Quantum Monte Carlo Simulation"}
\end{center}
\setcounter{equation}{0}
\setcounter{figure}{0}
\setcounter{table}{0}
\setcounter{page}{1}
\setcounter{section}{0}

\makeatletter
\renewcommand{\theequation}{S\arabic{equation}}
\renewcommand{\thefigure}{S\arabic{figure}}
\setcounter{secnumdepth}{3}

In Sec.~\ref{sec:SM1} of this Supplemental Material, we will provide a detailed derivation of the entanglement entropy (EE) as an exponential observable. We then discuss, in a general formalism, how to convert the expectation value of an exponential quantity into standard expectation value without exponential in the incremental algorithm, such that the complications and challenges in measuring exponential observables are eliminated and the computational complexity becomes identical to the measurement of regular physics observables in QMC.

In Secs.~\ref{sec:SM2}, we provide detailed derivation of the SCEE algorithm in the framework of incremental SWAP scheme. 

In Secs.~\ref{sec:SM3}, we demonstrate the SCEE is also an exponential observable.

In Sec.~\ref{sec:SM4}, we present additional data of SCEE by comparing the results from both incremental SWAP scheme and the non-equilibrium scheme, most importantly, we also demonstrate that the different cutting schemes of the entanglement boundaries, i.e. the standard and tilted cuts, reveal the same universal log-coefficient for $s(\pi/4)$ at the (2+1)D O(3) QCP.

In Sec.~\ref{sec:SM5}, we provide the QMC data for determining the precise position of the (2+1)D O(3) QCP for honeycomb Heisenberg bilayer from finite size analysis. 

\section{Exponential observable and its incremental solution}  
\label{sec:SM1}
The discussion of this section, follows closely with those in Refs.~\cite{zhangIntegral2023,daliaoControllable2023,zhouIncremental2024}.

In QMC simulations, the 2nd R\'enyi EE is calculated using the so-called replica trick, $S_{A}^{(2)}\equiv-\log(\Tr(\rho_{A}^2))$, it can be expressed according to configuration $\{s_1,s_2\}$ as
\begin{equation}
S_{A}^{(2)}=-\log \left(\frac{\sum_{s_1,s_2}P_{s_1}P_{s_2}\Tr \left(\rho_{A;s_1}\rho_{A;s_2}\right)}{\sum_{s_1,s_2}P_{s_1}P_{s_2}}\right),
\label{eq:eqS1}
\end{equation}
where $P_{s_i}$ is the Monte Carlo weight for configuration $\{s_i\}$, $\rho_{A;s_i}$ is the reduced density matrix and $\Tr(\rho_{A;s_1}\rho_{A;s_2})$ connects the two replicas. Depending on the implementation, $\Tr(\rho_{A;s_1}\rho_{A;s_2})$ is the SWAP operator in the projector QMC~\cite{hastingsMeasuring2010} and the non-equilibrium work done in the finite temperature non-equilibrium QMC~\cite{albaOutofequilibrium2017,demidioEntanglement2020} of quantum spin systems, and the determinant of Grover matrix~\cite{groverEntanglement2013,panStable2023} in fermion EE QMC simulations. With the understanding of exponential observable~\cite{zhangIntegral2023}, one views the 
\begin{eqnarray}
&&S_{A}^{(2)}=-\int_{0}^{1} dt \frac{\partial \log( f(t))}{\partial t} \nonumber\\
&&=-\int_{0}^{1} dt \frac{\sum P_{s_1}P_{s_2}\Tr(\rho_{A;s_1}\rho_{A;s_2})^{t} \log(\Tr(\rho_{A;s_1}\rho_{A;s_2}))}{\sum P_{s_1}P_{s_2}\Tr(\rho_{A;s_1}\rho_{A;s_2})^{t}}, \nonumber\\
\label{eq:eqS2}
\end{eqnarray}
with 
\begin{equation}
f(t) \equiv \sum P_{s_1}P_{s_2}\Tr(\rho_{A;s_1}\rho_{A;s_2})^{t},
\label{eq:eqS3}
\end{equation}
the entanglement partition function such that $f(1)=\sum P_{s_1}P_{s_2}\Tr(\rho_{A;s_1}\rho_{A;s_2})$ and $f(0)=\sum P_{s_1}P_{s_2}$. The nature of the exponential observable and its solution of incremental algorithm, i.e. $S^{(2)}_A=\hat{X}$, is  to implement 
\begin{equation}
 \log \langle  e^{\hat{X}} \rangle = \int_0^1 \mathrm{d}t \langle \hat{X}\rangle_t,
 \label{eq:eqS4}
 \end{equation}
where $\langle \ldots \rangle_t$ represents a modified average where the distribution probability is determined by an extra $t$ with the factor of $e^{t \hat{X}}$. Since the EE is generally an extensive quantity dominated by the area law, a direct simulation by using $P_{s_1}P_{s_2}$ as sampling weight and $\Tr(\rho_{A;s_1}\rho_{A;s_2})$ as observable according to Eq.~\eqref{eq:eqS1} (or the LHS of $\langle  e^{\hat{X}} \rangle$ in Eq.~\eqref{eq:eqS4}) will certainly give exponentially small values as $e^{-S_{A}^{(2)}} \sim e^{-a l_A}$. This is why the direct simulation according to Eq.~\eqref{eq:eqS1} is found to be unstable~\cite{hastingsMeasuring2010,assaad2014entanglement,broeckerRenyi2014,broeckerEntanglement2016,panStable2023}, i.e. one samples exponentially small observable which has exponentially large relative variances (fluctuations). What one needs, instead, is to sample according to the entanglement partition function $f(t)$ and compute the observable $\log(\Tr(\rho_{A;s_1}\rho_{A;s_2}))$, as in Eq.~\eqref{eq:eqS2} (or the RHS of Eq.~\eqref{eq:eqS4}), such that one converts the expectation value of an exponential quantity into standard expectation value without exponentials, and the computational complexity becomes identical to the measurement of regular physics observables. 

In fact, this is a very general practice and can be applied to all the exponential observable, not only the 2nd order R\'enyi EE but for the $n$th R\'enyi EE $S_{A}^{(n)}\equiv -\frac{1}{n-1}\log(\Tr(\rho_{A}^n))$ and free energy $F\equiv -\frac{1}{\beta}\log(Z) $ as shown in Ref.~\cite{zhangIntegral2023}.

In the main text, the incremental SWAP operator within the projector QMC for the EE computation, is one specific realization of such scheme in Eqs.~\eqref{eq:eqS2}, \eqref{eq:eqS3} and \eqref{eq:eqS4}. There, the incremental partition function $f(t)$ is
\begin{equation}
Z\left(k\right)=\sum_{C} W_{C} ( \text{SWAP}_A) ^{k/n},
\end{equation}
with $Z(0) = \sum_{C} W_{C}$ and  $Z({n}) = \sum_{C} W_{C} \text{SWAP}_A$ and the incremental process
\begin{equation}
 \langle \text{SWAP}_A\rangle=\frac{Z(1)}{Z(0)}\frac{Z(2)}{Z(1)}\cdots\frac{Z({k+1})}{Z(k)} \cdots\frac{Z({n})}{Z({n} - 1)},
\end{equation}
where integer $k$ denotes the $k$-th increment with $n$ total increments.

Subsequently, each increment is evaluated in parallel per PQMC as follows
\begin{equation}\label{eq:incre-slice}
\frac{Z\left({k+1}\right)}{Z\left(k\right)}=\frac{\sum\limits_{C} W_{C}  ( \text{SWAP}_A )^{k/ n} ( \text{SWAP}_A)^{1/n} }{\sum\limits_{C} W_{C} ( \text{SWAP}_A )^{k/n}}.
\end{equation}
The new sampling observable is $(\text{SWAP}_A )^{1/ n}$, it corresponds to the $\log(\Tr(\rho_{A;s_1}\rho_{A;s_2}))$ in Eq.~\eqref{eq:eqS2} which follows the normal distribution as shown in Ref.~\cite{zhouIncremental2024}. The new sampling weight is $W_{C} ( \text{SWAP}_A )^{k/ n}$ is the $f(t)$ in Eq.~\eqref{eq:eqS3} with $k/n \sim t$, to provide the sampling factor $e^{t\hat{X}}$ as the RHS in Eq.~\eqref{eq:eqS4}.

Therefore, the incremental SWAP operator is an application for exponential observable in QMC simulation, it allows us to evaluate the new observable $( \text{SWAP}_A )^{1/ n} $ for each piece of parallel incremental process with controlled statistical errors. As expected, all pieces of the incremental process should generate reliable samples that approximately follow a normal distribution and have a finite variance, as shown in Ref.~\cite{zhouIncremental2024}.

\section{The details for SCEE method}
\label{sec:SM2}
Here we demonstrate the SCEE method with explicit equations.
As mentioned in the main text, we have 
\begin{equation}
e^{-S_{A_1}^{(2)}} = \frac{ \sum\limits_{l,r} w_{l} w_{r}\left\langle V_{l}^0\left|(-H)^m \text{SWAP}_{A_1}   (-H)^m\right| V_r^0 \right\rangle}{  \sum\limits_{l,r} w_{l} w_{r} \left\langle  V_{l}^0 \left|(-H)^{2 m}\right|V_{r}^0 \right\rangle},
\end{equation}
and 
\begin{equation}
e^{-S_{A_2}^{(2)}} = \frac{ \sum\limits_{l,r} w_{l} w_{r}\left\langle V_{l}^0\left|(-H)^m \text{SWAP}_{A_2}   (-H)^m\right| V_r^0 \right\rangle}{  \sum\limits_{l,r} w_{l} w_{r} \left\langle  V_{l}^0 \left|(-H)^{2 m}\right|V_{r}^0 \right\rangle},
\end{equation}
We could obtain
\begin{equation}\label{eq:Seq5}
\begin{aligned}
e^{-S_s} &\equiv e^{-\left(S_{A_1}^{(2)}-S_{A_2}^{(2)}\right)} \\
&=\frac{ \sum\limits_{l,r} w_{l} w_{r}\left\langle V_{l}^0\left|(-H)^m \text{SWAP}_{A_1}   (-H)^m\right| V_r^0 \right\rangle}{  \sum\limits_{l,r} w_{l} w_{r} \left\langle  V_{l}^0 \left|(-H)^{2 m}\right|V_{r}^0 \right\rangle} / \frac{ \sum\limits_{l,r} w_{l} w_{r}\left\langle V_{l}^0\left|(-H)^m \text{SWAP}_{A_2}   (-H)^m\right| V_r^0 \right\rangle}{  \sum\limits_{l,r} w_{l} w_{r} \left\langle  V_{l}^0 \left|(-H)^{2 m}\right|V_{r}^0 \right\rangle} \\
&= \frac{ \sum\limits_{l,r} w_{l} w_{r}\left\langle V_{l}^0\left|(-H)^m \text{SWAP}_{A_1}   (-H)^m\right| V_r^0 \right\rangle}{  \sum\limits_{l,r} w_{l} w_{r} \left\langle  V_{l}^0 \left|(-H)^m \text{SWAP}_{A_2}   (-H)^m\right|V_{r}^0 \right\rangle} \\
&= \frac{\sum\limits_{l,r} \bm{W}_{l,r,A_2} \bm{O}_{l,r,A_1,A_2}}{\sum\limits_{l,r} \bm{W}_{l,r,A_2}},
\end{aligned}
\end{equation}
where 
\begin{equation}
\bm{W}_{l,r,A_2} = w_{l} w_{r}\left\langle V_{l}^0\left|(-H)^m \text{SWAP}_{A_2}   (-H)^m\right| V_r^0 \right\rangle
\end{equation} 
and  
\begin{equation}
\bm{O}_{l,r,A_1,A_2} = \frac{\left\langle V_{l}^0\left|(-H)^m \text{SWAP}_{A_1}   (-H)^m\right| V_r^0 \right\rangle }{\left\langle V_{l}^0\left|(-H)^m \text{SWAP}_{A_2}   (-H)^m\right| V_r^0 \right\rangle
}. 
\end{equation}

\section{$S_s$ is an exponential observable}
\label{sec:SM3}
\begin{figure}[htp!]
\includegraphics[width=0.6\columnwidth]{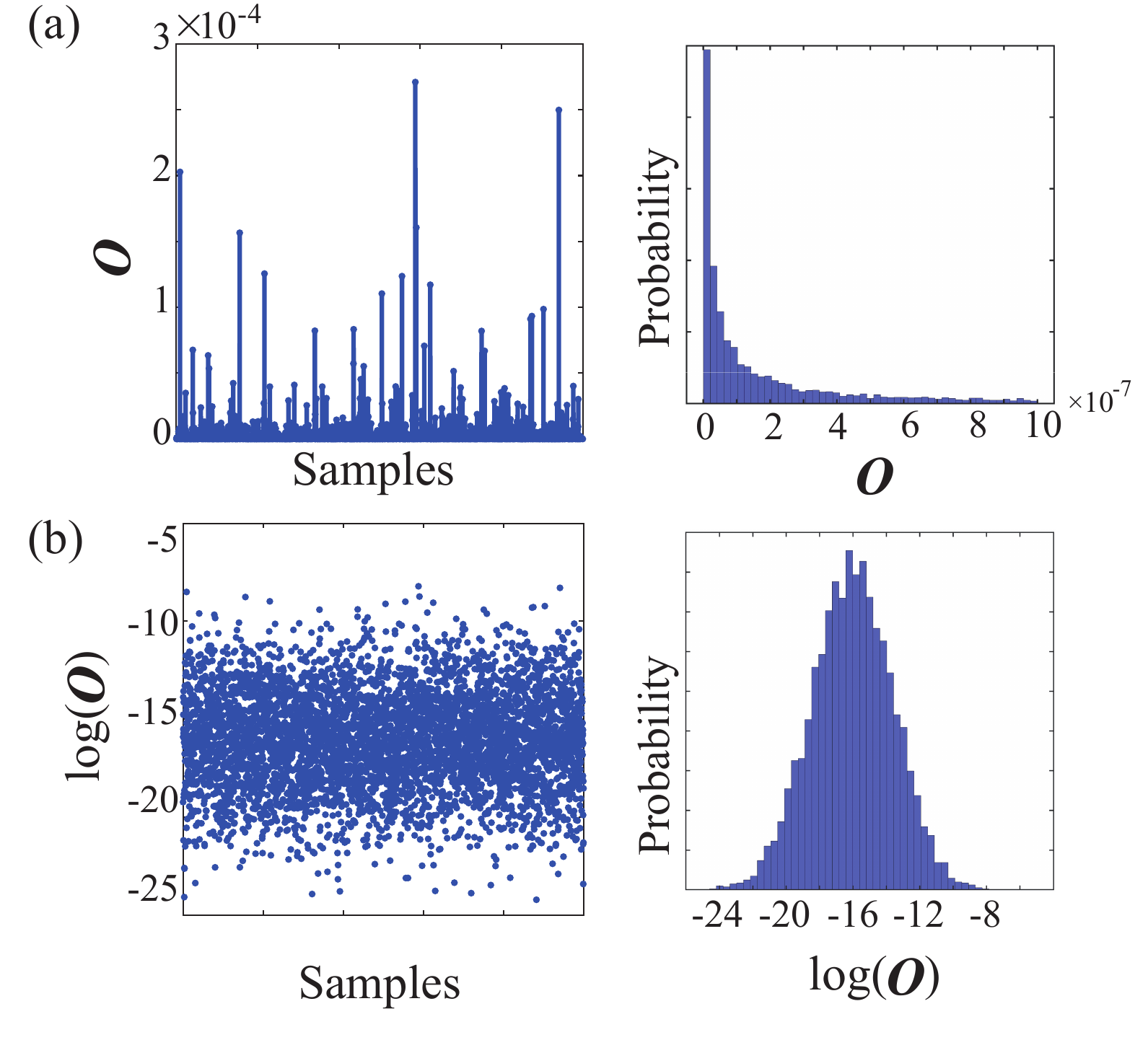}
\caption{The distribution of  $\bm{O}$ (a) and $\ln \bm{O}$ (b) for the bilayer Heisenberg model on standard square lattice at $g_c$ when $L=20$.}
\label{fig:S1}
\end{figure}

As depicted in Fig.~\ref{fig:S1}(a), the distribution of $\bm{O}$ follows a log-normal distribution when we simply use Eq.~\eqref{eq:Seq5} to compute $S_s$ for the bilayer Heisenberg model at $g_c=2.522$ on standard square lattice. The distribution of $\ln \bm{O}$ conforms well to the normal distribution, as illustrated in Fig.~\ref{fig:S1}(b). These observations indicate that, similar to the exponential observable 2nd R\'enyi EE $S_A^{(2)}$, $S_s$ is also an exponential observable. 
In other words, we need to employ the incremental SWAP operator method to calculate $S_s$ rather than simply simulating it according to Eq.~\eqref{eq:Seq5}.

\section{Additional EE data from SCEE}  
\label{sec:SM4}
\begin{figure}[htp!]
\includegraphics[width=0.8\columnwidth]{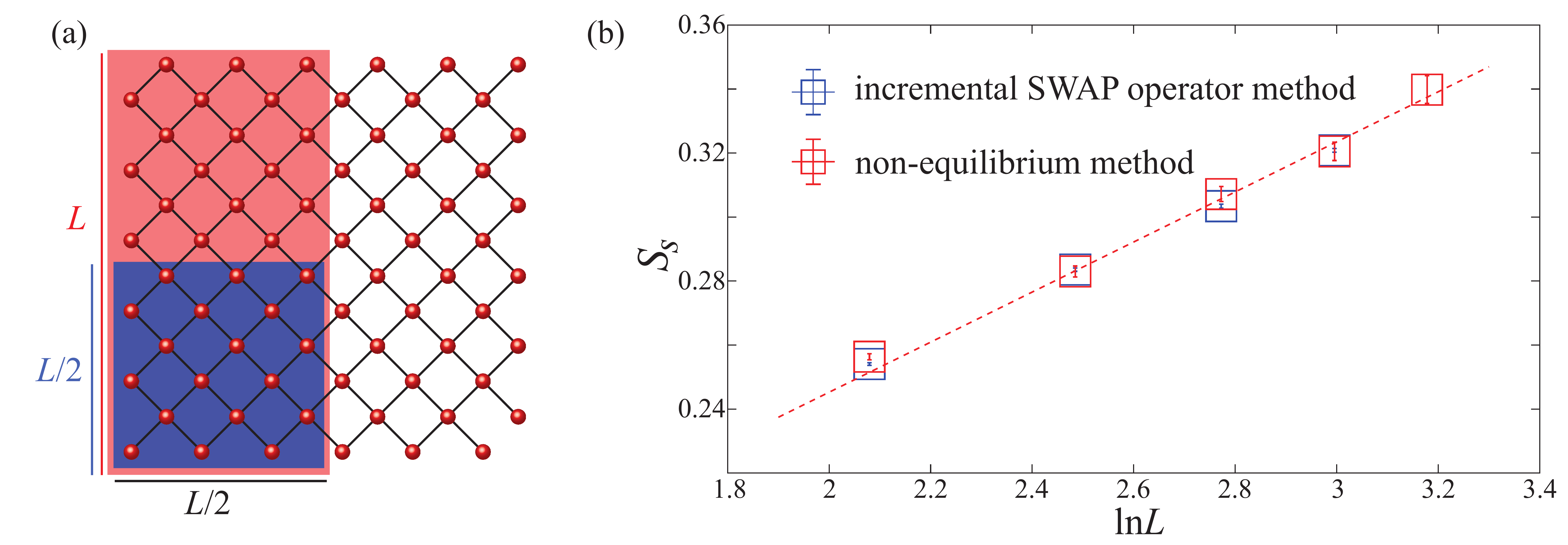}
\caption{(a) Depiction of the tilted square lattice and the entanglement region chosen for the SCEE $S_s$. (b) Comparison of the $S_s$ for the 2D bilayer Heisenberg model at the QCP obtained using incremental SWAP operators and non-equilibrium methods. The results of the $S_s$ obtained with the two different methods are comparable within error bars. The extracted $s(\pi/2)=0.078(5)$ is based on data points obtained with the non-equilibrium method for $L=12$ to $24$.}

\label{fig:tilt}

\end{figure}

We have also calculated the SCEE of the 2D bilayer Heisenberg model at the QCP for a tilted square lattice, as depicted in Fig.~\ref{fig:tilt} (a).
Recent study has reported that the universal term of the EE from sharp corners differs in J-Q models at the transition point for standard and tilted square lattices. In this work, we demonstrate that the universal term of the EE from four $\pi/2$ corners at the (2+1)D QCP remains consistent between standard and tilted square lattices, as shown in Fig.~\ref{fig:tilt} (b).
We have extracted a universal coefficient $s(\pi/2)=0.078(5)$ for the 2D bilayer Heisenberg model on a tilted square lattice at $g=2.5220$, while $s(\pi/2)=0.080(3)$ for standard square lattice.

\section{QMC determination of the QCP for 2D bilayer Heisenberg model in honeycomb lattice}
\label{sec:SM5}
Previous theoretical and numerical studies have computed the quantum critical point, $g_c\equiv J_\perp/J$ of the 2D bilayer Heisenberg model in honeycomb lattice~\cite{oitmaaGround2012,Ganesh2011Neeltodimer}. We perform an independent benchmark and present the data from large-scale quantum Monte Carlo simulation in this section. Our determined $g_c=1.644(1)$ is highly consistent with a previous study which reports a $g_c=1.645(1)$.

\begin{figure}[htp!]
\includegraphics[width=0.95\columnwidth]{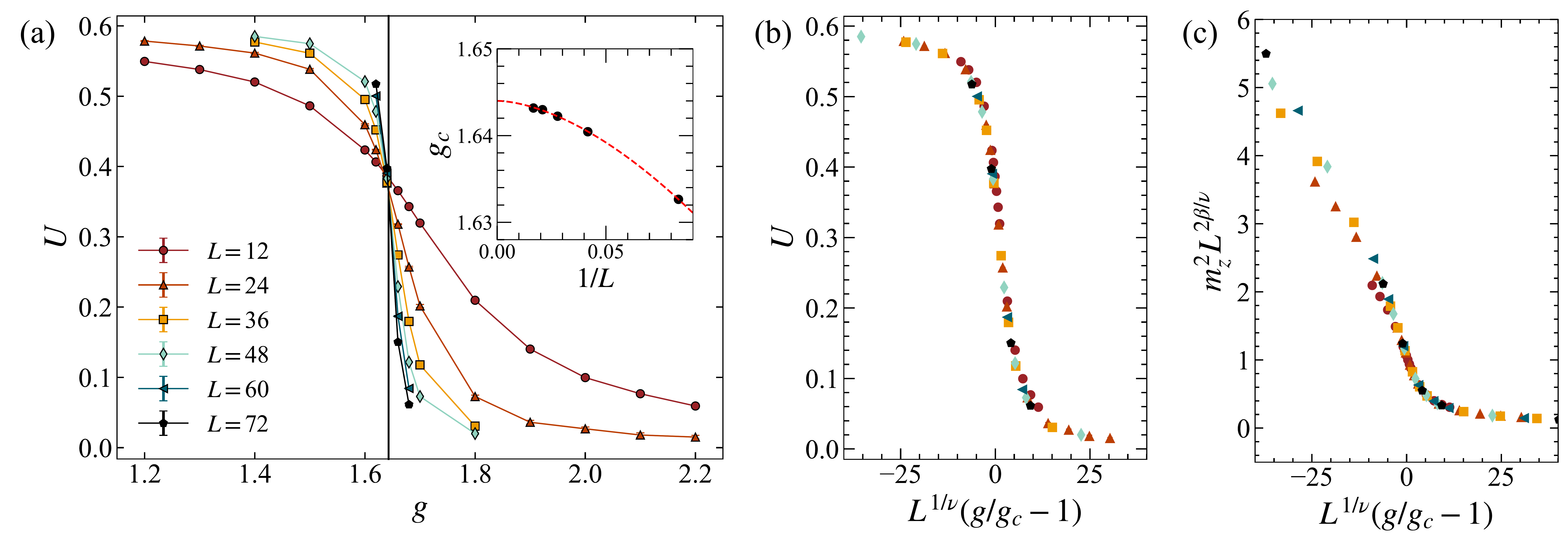}
\caption{\textbf{Binder ratio and order parameter for 2D antiferromagnetic honeycomb bilayer Heisenberg model.} (a) The critical points $g_c\equiv J_\perp/J$ is determined by the binder ratio crossing point of two consecutive system sizes. Binder ratio curves of various system sizes cross at $g=g_c$, shown as the vertical black line. The inset shows the system size dependence of the crossing point for two consecutive system sizes, which converges to $g_c=1.644(1)$ as depicted by the red line. (b) and (c) demonstrate the data collapse for the binder ratio and order parameter, respectively, using the critical exponents of (2+1)D O(3) universality class.}
\label{fig:qcp_honey}
\end{figure}

The critical point $g_c$ is determined from the crossing of the N\'eel order Binder ratios,  $$U = \frac{3}{2}(1-\frac{1}{3}\frac{\langle m^4_z\rangle }{\langle m^2_z\rangle ^2}).$$ 
Fig.~\ref{fig:qcp_honey}(a) shows the crossing of binder ratio curves for various system sizes. The inset illustrates that the crossing points converge to a finite value at the thermal dynamic limit, extrapolating a $g_c=1.644(1)$. In addition, as shown in Fig.~\ref{fig:qcp_honey} (b) and (c), we obtain satisfactory data collapse results for both binder ratio $U$ and order parameter $m_z^2$ using critical exponents of the (2+1)D O(3) universality class, i.e., $\nu=0.7112$, $\beta=0.3689$~\cite{campostrini2002criticalexponents,wang2006bilayer}, indicating the transition belongs to the (2+1)d O(3) universality class.

\section{The raw data of $S_s$ for 2D bilayer Heisenberg model}  
\begin{figure}[htp]
\centering
\includegraphics[width=0.7\columnwidth]{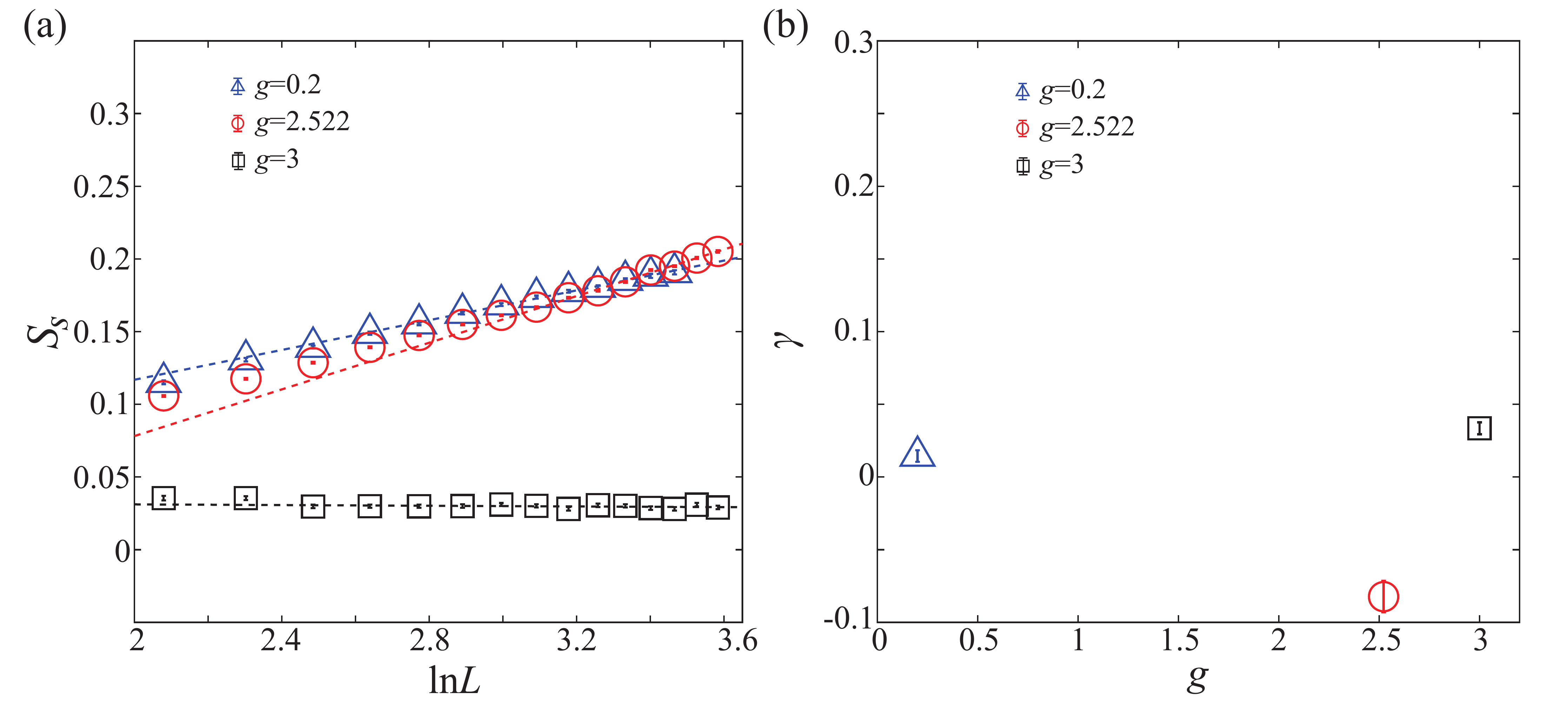}
\caption{\textbf{The $S_s$ for 2D bilayer Heisenberg model on square lattice with four $\pi/2$ corners.}}
\label{fig:S2}
\end{figure}

\begin{figure}[htp]
\centering
\includegraphics[width=0.7\columnwidth]{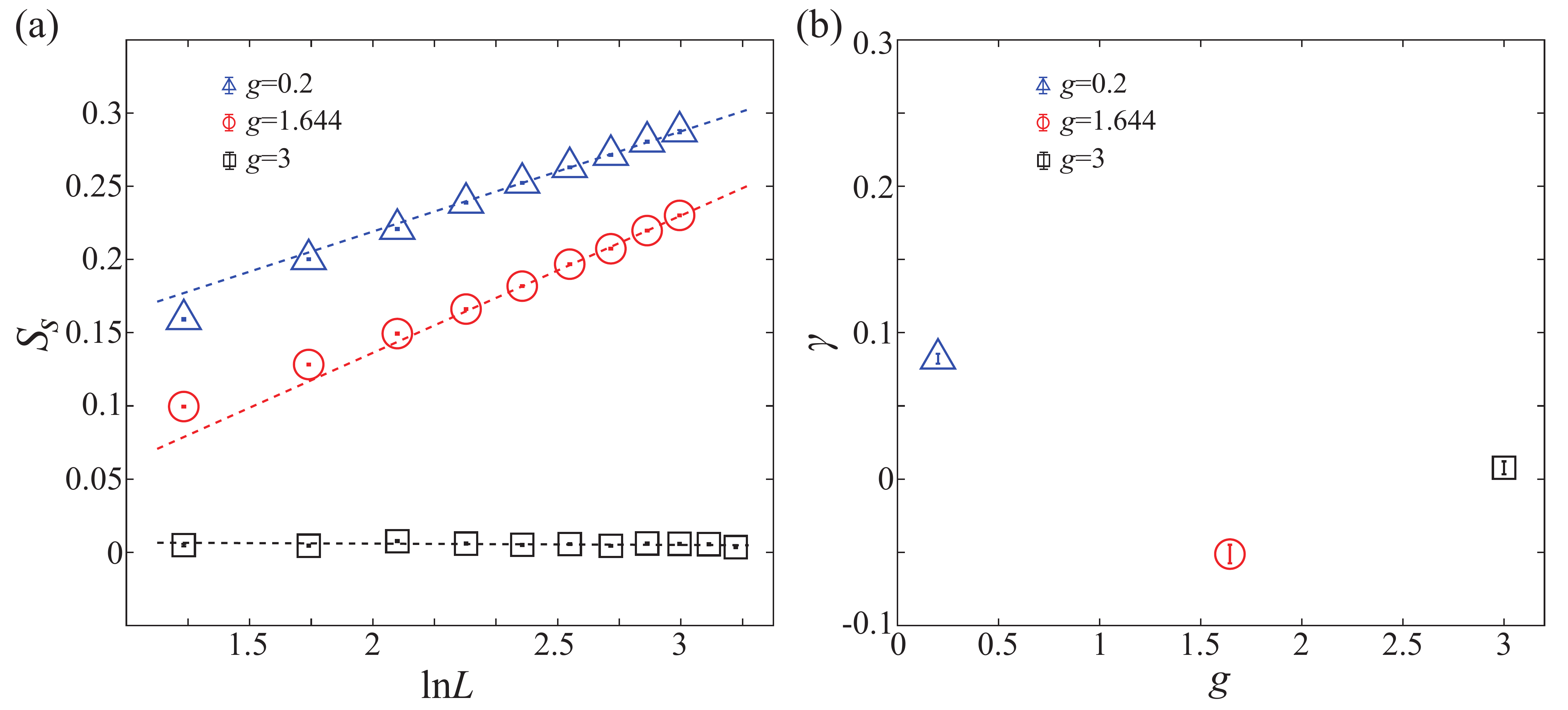}
\caption{\textbf{The $S_s$ for 2D bilayer Heisenberg model on square lattice with four $\pi/2$ corners.}}
\label{fig:S3}
\end{figure}

Using the SCEE method within the incremental SWAP framework, we obtained precise $S_s$ values with very small error bars $e_{S_s}$, as shown in Fig.~\ref{fig:S2}(a) and Fig.~\ref{fig:S3}(a).
We then used equation, $S_s = s\ln L + \gamma$, to fit the corner coefficient $s$ with error $e_s$ and the constant term $\gamma$ with error $e_\gamma$.

The primary reason we chose to plot $S_s-\gamma$ instead of $S_s$ was to enhance the clarity of data presentation. As one can see in Fig.~\ref{fig:S2}(a), the raw $S_s$ data points are overlapping, making it difficult to discern individual trends. By subtracting $\gamma$, we were able to separate the data points, allowing for a clearer visualization of the behavior across different system sizes and parameters.

\begin{figure}[htp]
\centering
\includegraphics[width=0.9\columnwidth]{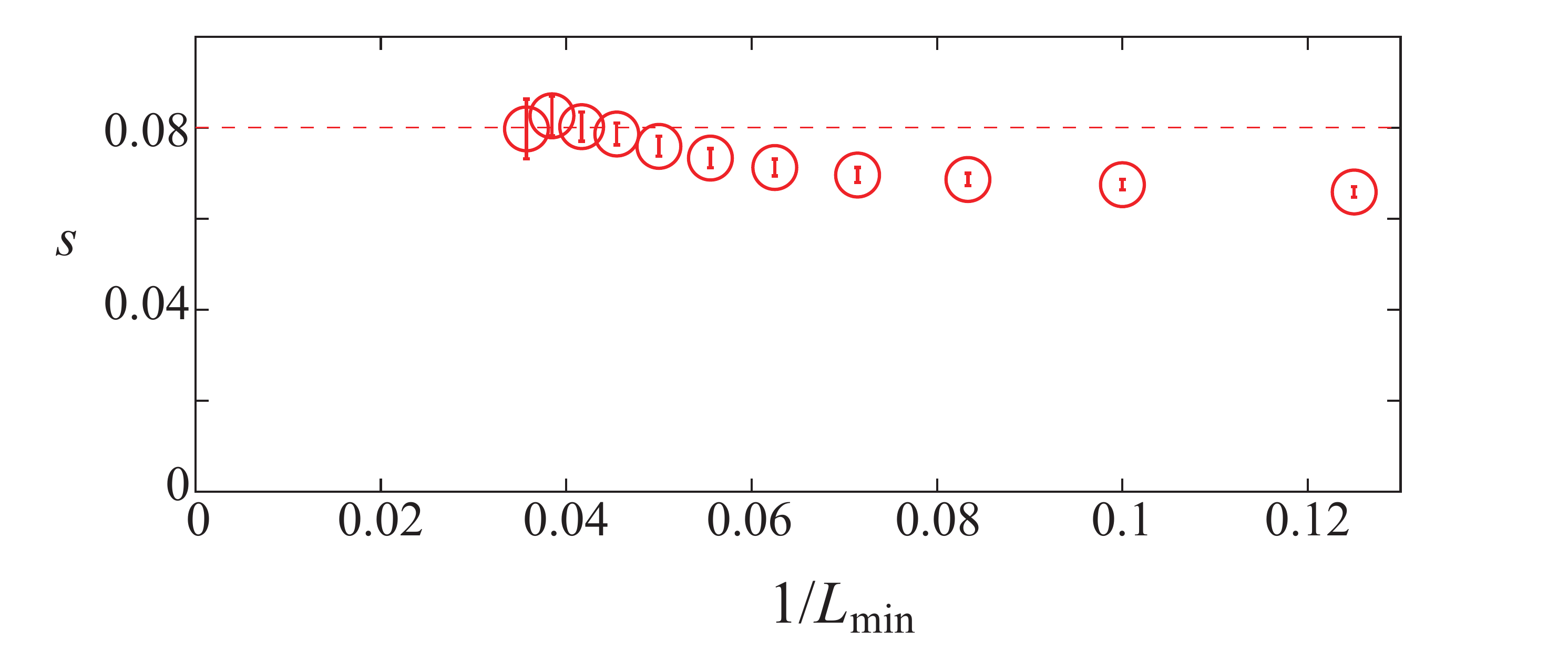}
\caption{\textbf{The fitted corner coefficients using a fit window $L_{\text{min}}$ to $L_{\text{max}}=36$.}}
\label{fig:s4}
\end{figure}

Furthermore, as smaller system sizes are subject to significant finite-size effects, necessitating their exclusion to accurately extract the universal corner coefficient.
To further demonstrate the validity of our approach, we have plotted the $1/L_{\text{min}}$ extrapolation of the corner coefficient at the QCP $g=2.522$ with the fitting window $[L_{\text{min}}, L_{\text{max}}=36]$.
As shown in Fig.~\ref{fig:s4}, the corner coefficient $s$ at the QCP converges around $L_{min}=24$. 

\end{document}